# The relationship among research productivity, research collaboration, and their determinants[1]

Giovanni Abramo

*Laboratory for Studies in Research Evaluation*
*at the Institute for System Analysis and Computer Science (IASI-CNR)*
*National Research Council of Italy*
ADDRESS: Istituto di Analisi dei Sistemi e Informatica, Consiglio Nazionale delle Ricerche, Via dei Taurini 19, 00185 Roma - ITALY
tel. +39 06 7716417, fax +39 06 7716461, giovanni.abramo@uniroma2.it

Ciriaco Andrea D'Angelo

*University of Rome "Tor Vergata" - Italy and*
*Laboratory for Studies in Research Evaluation (IASI-CNR)*
ADDRESS: Dipartimento di Ingegneria dell'Impresa, Università degli Studi di Roma "Tor Vergata", Via del Politecnico 1, 00133 Roma - ITALY
tel. and fax +39 06 72597362, dangelo@dii.uniroma2.it

Gianluca Murgia

*University of Siena - Italy*
ADDRESS: Dipartimento di Ingegneria dell'Informazione e Scienze Matematiche, Università degli Studi di Siena, Via Roma 56, 53100 Siena - ITALY
tel. and fax +39 0577 1916386, murgia@dii.unisi.it

**Abstract**
This work provides an in-depth analysis of the relation between the different types of collaboration and research productivity, showing how both are influenced by some personal and organizational variables. By applying different cross-lagged panel models, we are able to analyze the relationship among research productivity, collaboration and their determinants. In particular, we show that only collaboration at intramural and domestic level has a positive effect on research productivity. Differently, all the forms of collaboration are positively affected by research productivity. The results can favor the reexamination of the theories related to these issues, and inform policies that would be more suited to their management.

**Keywords**
*Bibliometrics; research productivity; cross lagged panel model; direct and indirect effects; university.*

# 1. Introduction

The scientific literature on the determinants of a researcher's performance (Costas, Van Leeuwen & Bordons, 2010; Gonzalez-Brambila & Veloso, 2007; Harris & Kaine, 1994; Cole & Zuckerman, 1984) has shown how this depends on numerous personal and organizational variables. These variables influence the level of competencies, the resources and time available, and the individual's motivation and reputation, which are at the basis of the performance. The continuing decline in the share of single-authored publications (Uddin, Hossain, Abbasi & Rasmussen, 2012) has often been associated with the growth in research performance (Gonzalez-Brambila, Veloso & Krackhardt, 2013; He, Geng & Campbell-Hunt, 2009). In particular, as research collaboration increases, the number of publications (Ductor, 2015; Lee & Bozeman, 2005) and citations also increases (Bidault & Hildebrand, 2014; Li, Liao & Yen, 2013).

The link between research collaboration and performance would at this point seem accepted in the literature (He, Geng & Campbell-Hunt, 2009; Lee & Bozeman, 2005), however in fact there has not been full clarification of the causal nexus between collaboration and research performance. First of all, only few papers have specifically analyzed the different forms of collaborations (intra-university, domestic, and international), which feature notably different efficacies and costs (He, Geng & Campbell-Hunt, 2009; Smeby & Try, 2005). Secondly, even fewer papers have tested the impact of research performance on the ability to activate collaborations, while most of the literature has considered only the opposite causal mechanism (He, Geng & Campbell-Hunt, 2009; Landry & Amara, 1998). Finally, although in a different manner, both research collaboration and performance are influenced by the same personal and organizational variables (Lee & Bozeman, 2005). Besides, some of these variables, firstly academic rank, could in turn be influenced by research performance and collaboration (Lissoni, Mairesse, Montobbio & Pezzoni, 2011; Pezzoni, Sterzi & Lissoni, 2012). The presence of such opposite and interrelated causal mechanism requires the use of statistical methods that are able to deal with the endogeneity among the variables under analysis.

In this study we adopt a structural equation modelling approach to estimate different cross lagged panel models. Although these models do not provide a definitive answer to the causal relationship among research productivity, research collaboration, and some of their determinants, such as gender, cohort, and academic rank, they allow us to measure the strength of the different relationships among these variables. We can then evaluate the impact of different forms of propensity to collaborate on research productivity, and vice versa. Besides, we are also able to evaluate how the determinants affect research collaboration and productivity in indirect way, thanks to the mediation of other variables.

Compared to the preceding studies on this topic (He, Geng & Campbell-Hunt, 2009; Lee & Bozeman, 2005), the present work is distinguished also for the breadth and exhaustiveness of the field of observation, that is a large share of the population of the professors of Italian universities in the areas of the sciences and economics (16,823 in all). Another notable element of the present paper is in the indicator of research performance, given that we measure research productivity by means of the (fractional) total impact, meaning the sum of the field-normalized citations received by the publications of each professor (Abramo & D'Angelo, 2014).

The next section of the paper draws from an analysis of the literature to indicate an expected framework of links between collaboration, academic rank, research productivity and the other personal and organizational variables under consideration. Section 3 describes the dataset and the methodology used. Section 4 presents the empirical results, while the implications of the findings are discussed in Section 5.

## 2. Literature review

To understand how the research productivity of a scientist can be influenced by their research collaboration, it is appropriate to begin from the factors that determine high research performance, measurable in terms of the number of publications and their impact (Figure 1).

Above all, the advancement of scientific knowledge demands that the researcher be equipped with the appropriate competencies, beginning from knowledge about the problem under analysis, which permits the individual to carry out original and appropriate analyses, up to the necessary competencies in methodologies and reporting the findings in publication. The increasing multidisciplinarity and complexity that characterizes current scientific research often results in contexts where a single scientist does not possess all the necessary competencies for the achievement of scientific advancement (Beaver, 2001; Katz & Martin, 1997). Collaboration permits overcoming these shortcomings by means of involving scientists who are specialized in the missing competencies. Moreover, collaboration facilitates the generation and selection of original ideas, thanks to the synergies that can be obtained from scientists with complementary backgrounds, or even from different disciplines (Rigby & Edler, 2005; Katz & Martin, 1997). This process is especially favored by international collaborations, because they involve scientists gifted with complementary competencies, and mindsets that are often sharply differentiated (Burt, 1992). Over a longer time horizon, collaboration permits overcoming the individual's gaps in competencies, through the activation of learning processes including learning of tacit knowledge (He, Geng & Campbell-Hunt, 2009; Beaver, 2001). The involvement of multiple authors also permit a more efficient use of time and limits the need to the resort to external advisors, for example for third party checking of research processes and outcomes (Barnett, Ault & Kaserman, 1988).

In many cases, the achievement of scientific projects requires not only competencies, but also equipment and various other resources. Collaboration can ensure the access to unique or costly resources, through the involvement of research groups already equipped with these assets (Beaver & Rosen, 1978) or through developing multiple research projects, each one featuring an adequate critical mass of scientists (Beaver, 2001). It is no accident that the greatest resort to collaboration, even at international level, occurs in the "big science" disciplines, where there is greater need to access unique equipment or data (Katz and Martin, 1997).

Time is often perceived as the most critical ingredient for the achievement of scientific advancement, in part because of the other personal and academic activities in which the scientist is engaged. Collaboration can permit the more efficient use of time, because it permits the division of labor between the different team members (Barnett, Ault & Kaserman, 1988). The division of labor often permits the team members to work in parallel on various parts of a project, reducing the time for its accomplishment (Beaver, 2001). Concerning the preparation of the publications attesting the achieved results, collaboration permits the individual authors to distribute this task more broadly, which among other benefits can reduce the risk of rejection by editors of scientific journals (Beaver, 2001; Barnett, Ault & Kaserman, 1988). Clearly, collaboration can also bring about an increase in the time that the individual scientist dedicates to preparing the publication, especially when coordination with the other authors is inefficacious (Lee & Bozeman, 2005; Landry, Traore & Godin, 1996) or when the various authors are slowed by other work (Ductor, 2015).

For a scientist to spend their available time in the execution of research projects and the subsequent publication of the results, rather than on other academic and personal activities, it is also necessary that the individual be sufficiently motivated. Collaboration can favor a scientist's motivations to achieve research and publications, because it offers a manner to overcome intellectual isolation (Beaver & Rosen, 1978), sharing their enthusiasm for the problem with other scientists while satisfying their own curiosity. The scientist also feels a responsibility towards his/her collaborators, because their work also bears on their research output (Beaver, 2001). In addition, collaboration permits the activation of pleasurable social interaction, which can improve the climate within the research group (Medoff, 2003).

The levels of competencies, resources, time and motivation strongly influence the mass of research results that an individual researcher can produce, and as a result also influences the number of publications authored. Laband (1985) and Piette and Ross (1992) illustrate how the reputation of the authors themselves influences the success rate of submission to scientific journals. The quality of the publication influences the number of citations received, but the numbers are also influenced by the author's reputation. With the increasing reputation of the author, there is an increase in the propensity to cite their publications, on the part of the other scientists. The reputation of a scientist can also be favored by collaboration, since it facilitates the diffusion of his/her discoveries (Katz & Martin, 1997). It is above all collaborations with prestigious scientists that guarantee greater visibility to the other authors (Beaver, 2001). Moreover, international collaborations favor an increase in the number of citations received thanks to the joint authors' more extended network of contacts (Schmoch & Schubert, 2008; Goldfinch, Dale & DeRouen Jr, 2003).

*Figure 1: A systematic view of the determinants of research productivity*

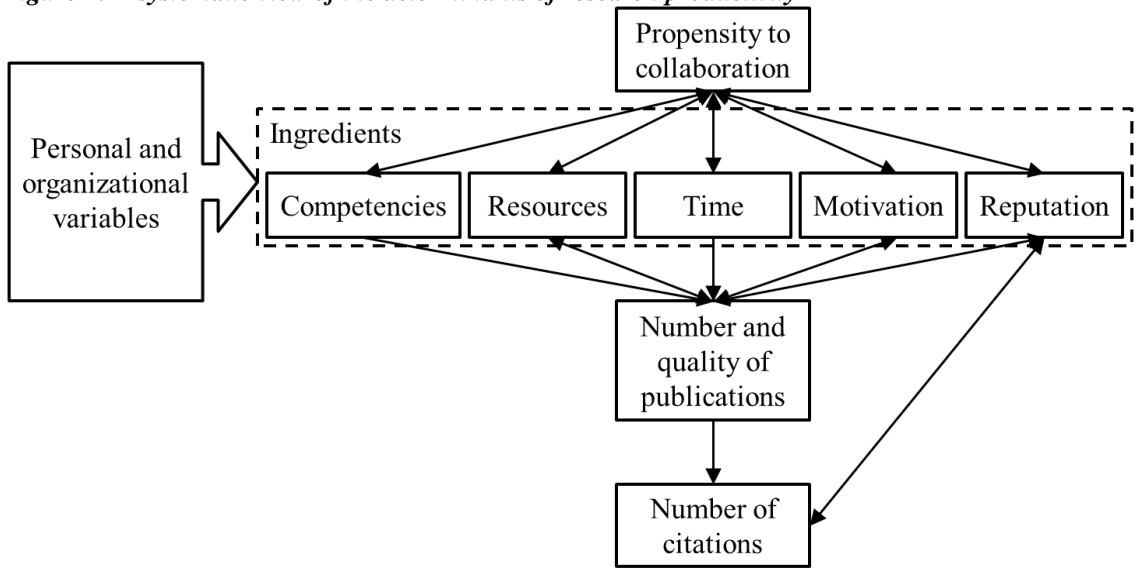

The advantages to be had by means of collaboration are confirmed by empirical findings, which have evidenced a positive impact on research performance (Ductor, 2015; McFadyen & Cannella, 2004). However, sometimes a scientist does not develop collaborations for the primary objective of achieving knowledge advancements, but to improve the relational and political dimensions of their own social capital (Pezzoni, Sterzi & Lissoni, 2012; Ynalvez & Shrum, 2011). This would at a maximum have only an indirect impact on their research productivity. Moreover, a scientist capable of publishing alone already possesses all the ingredients necessary for the achievement of publications, at least on certain themes, a situation which the individual could then use to select and organize a limited number of collaborations in the most advantageous manner possible (Medoff, 2003; Hollis, 2001). For this reason, a scientist with a lesser propensity to collaborate could generally be characterized by a greater research productivity.

Finally, it is possible that the time and the cost necessary for the coordination of a collaboration network could be such as to reduce the research productivity of a scientist (Ynalvez & Shrum, 2011; Landry, Traore & Godin, 1996). The absence of effective and economical transport and communications systems can hinder the possibility of human direct interaction and transmission of tacit knowledge and the creation of the climate of trust and cooperation necessary for the achievement of joint research projects and the subsequent coauthored publications (Ynalvez & Shrum, 2011). In many contexts, technological evolution has favored overcoming the limits of international collaborations, which in fact have significant and positive impacts on research productivity, particularly when the quality of the publications is considered (Lissoni, Mairesse, Montobbio & Pezzoni, 2011; He, Geng & Campbell-Hunt, 2009; Smeby & Try, 2005). Still, the cost of coordinating the collaborations at the domestic level, and above all at the intra-university level, remains lower, and it is therefore plausible that these types of collaborations can impact significantly on the number of publications achieved by a researcher (He, Geng & Campbell-Hunt, 2009; Smeby & Try, 2005). On the other hand, the relatively low costs connected to such collaborations could explain the qualitative gap with respect to the publications resulting from international collaborations. Indeed the greater cost of international collaborations induces a selection of coauthors that is oriented towards the maximization of productivity, while the collaborations at domestic and intra-university level could also be motivated by different factors (Shin & Cummings, 2010; Schmoch & Schubert, 2008). In particular, domestic and intra-university collaborations could be a means to increase a scientist's political social capital within his or her university and discipline. The political social capital is a critical factor in determining career advancement within the Italian academic system (Pezzoni, Sterzi & Lissoni, 2012).

Eventually, contrasting results in the relationship between collaboration and research performance could also be due to the different impacts of the various forms of collaboration, defined in terms of the spatial distance between the coauthors (at the intra-university, domestic and international level). These different forms of collaboration permit access to competencies and resources of differing character, and impact differently on the motivation and reputation of the scientist, as well as on the means of coordination. As we will describe in Section 3, our study evaluates research productivity taking into account both the quantity and the quality of publications. Therefore, we formulate the following hypothesis:

*H1: The propensity to collaborate generally has a positive impact on research productivity. This positive effect is higher for international collaborations than for those at the domestic or intra-university level.*

The development of scientific and technical human capital often comes about in a contextual manner with respect to the growth of social capital (Bozeman & Corley, 2004). In fact, the assets of individual competencies, resources, time, motivation and reputation do not bear only on research performance (for example, on the number of publications and citations of a scientist), but also on the person's propensity to collaborate, seeing as they permit the attraction and management of potential collaborators in a more effective manner. Abramo, D'Angelo and Solazzi (2011), and Kato and Ando (2013) argue that research productivity has a positive impact on international collaboration, while He, Geng and Campbell-Hunt (2009) shows that only the opposite causal relationship is statistically supported. Given these results, we formulate the following hypothesis:

*H2: The research productivity generally impacts in positive manner on the propensity to collaborate. This positive effect is higher on collaborations at the international level than those at the domestic or intra-university level.*

The understanding of the relation between collaboration and research performance cannot be complete if we do not account for the personal and organization variables that influence both, although the path of these influences is in different manner. In the literature, only Lee and Bozeman (2005) simultaneously evaluate the effect of these variables on research performance and collaboration, by using a mediation analysis based on two-stage least squares (2SLS) method. This method does not permit the comparison of the sizes of the direct and indirect effects, by means of an evaluation of the portion of the total effect due to the indirect effect (Iacobucci, 2012). In order to overcome this limitation, in the present study, we adopt a mediation analysis based on a Structural Equation Modelling, as described in section 3. In particular, in this work, we will consider researcher's cohort, gender, and academic rank, evaluating their direct and indirect effect on research collaboration and productivity.

In the analysis of the impact of academic rank, it is necessary to highlight, first of all, that research performance and propensity to collaborate have generally a positive impact on promotion. In fact, a certain level of research productivity and collaboration, especially at international level, is often required in an ever more meritocratic and globalized university system (Lissoni, Mairesse, Montobbio & Pezzoni, 2011; Pezzoni, Sterzi & Lissoni, 2012; van Rijnsoever, Hessels & Vandeberg, 2008). So, given a context of promotion based on research productivity and collaboration, the academics of higher rank tend to gain greater benefit due to the emergence of cumulative advantage, through which they can maintain and grow their competencies, resources and reputation with less effort, compared to what is necessary for their lower ranked colleagues (Cole & Cole, 1973; Merton, 1968). Indeed, various studies have shown that academic rank positively influences research productivity (Mishra & Smyth, 2013; Abramo, D'Angelo, & Di Costa, 2011; Kelchtermans & Veugelers, 2011). The greater research performance of the higher academic ranks can be in part due to their greater propensity to collaborate, because academics of higher rank generally have larger collaboration networks compared to those of lower rank (Abramo, D'Angelo, & Di Costa, 2011; Bozeman & Gaughan, 2011). This is due both to their duties in mentoring and project management, which are primarily assigned to the higher rank academics (Martin-Sempere, Garzon-Garcia & Rey-Rocha, 2008; Bayer & Smart, 1991), and to their greater capacity for attraction of potential collaborators, due to superior assets of resources, competencies and reputation (Abramo, D'Angelo & Murgia, 2014; Tien & Blackburn, 1996). Finally, academics of higher rank often take a greater advantage to the emergence of phenomena of "gift" authorship, above all from mentoring relationships (Bayer & Smart, 1991). Comparing academic ranks, full professors show the greatest propensity to collaborate at the international level (Abramo, D'Angelo & Murgia, 2014; Bozeman & Corley, 2004; Melkers & Kiopa, 2010), while the assistant professors at the intramural level, and the associate professors at the domestic level (Abramo, D'Angelo & Murgia, 2014). Academic rank could have also a negative impact on research productivity because the academics of higher rank have less time to devote to research activity, due to their greater teaching and service duties (Mishra & Smyth, 2013). Besides, the growth of research productivity with advancing academic rank could in some cases be hindered by lesser motivation to do research, experienced among full professors in the absence of possibilities for further promotion (Abramo, D'Angelo & Murgia, 2016; Kyvik & Olsen, 2008). Still, full professors are induced to maintain and improve their productivity in order to protect the prestige obtained through their previous performance (Kelchtermans & Veugelers, 2011). For these reasons, we formulate the following hypothesis:

*H3: The higher the academic rank the greater the positive impact on research performance and collaboration, especially at the international level.*

The continuous progress of knowledge could suggest that the younger cohorts of academics are better educated and more capable in research activities than the older ones (Levin & Stephan, 1991). Furthermore, increasingly competitive higher education systems are more demanding in terms of research performance for new entrants. For example, in the Italian university system, the growing emphasis on publications as a criterion for hiring could have strengthened the performance gap between older and younger cohorts. In a number of disciplines, such increase in productivity comes along with an increasing number of co-authors in the publications' bylines, and it is quite rare that a scientist, especially if belonging to the younger cohorts, produce sole-authored paper. Significant differences in the propensity to collaborate are observed as well, among the various cohorts of academics, because of a lesser international opening for the older cohorts (Kyvik & Olsen, 2008). For these reasons, we formulate the following hypothesis:

*H4: The younger the cohort the greater research performance and intensity of collaboration, especially at the international level.*

In the same way, gender also influences research productivity. The lesser productivity of female researchers has been established in tens of studies of diverse disciplines and countries (Larivière, Vignola-Gagné, Villeneuve, Gelinas & Gingras, 2011; Abramo, D'Angelo & Caprasecca, 2009; Mauleón & Bordons, 2006). The gap seems to be most visible in the early career stages (Xie & Shauman, 1998; Cole & Zuckerman, 1984; Kyvik & Teigen, 1996). This is above all due to the lesser availability of time to dedicate to research, due to women's acceptance of family responsibilities (Kyvik & Teigen, 1996). Moreover, above all in the disciplines dominated by an "old boy network", the female scientist can have difficulty in obtaining the necessary resources for their research, and their motivation is not strengthened by adequate social support (Etzkowitz, Kemelgor & Uzzi, 2000). These two factors impact on the female researchers' propensity to collaborate, and in fact they tend to develop collaboration networks that are less international (Bozeman & Corley, 2004; Abramo,

D'Angelo & Murgia, 2013b). Therefore, we formulate the following hypothesis:

*H5: Male academics show greater research performance and intensity of collaboration than females, especially at the international level.*

## 3. Methods and data

*3.1 Data sources and field of observation*

A brief presentation of the characteristics of the Italian university system assists in interpreting our research results. In keeping with the Humboldtian model, there are no "teaching-only" universities in Italy, as all professors are required to carry out both research and teaching. The academics are hired in the manner of tenured civil servants, and their salaries and contractual terms of work are fixed at the national level. In the Italian higher education system each professor is classified in one and only one research field, named Scientific Disciplinary Sector (SDS), of which there are 370,[2] grouped in 14 disciplines, named University Disciplinary Areas (UDAs).

Our analysis is limited to the fields (SDS) where bibliometrics can be applied. As a criterion, we chose to analyze only those SDSs (198 in all) where in the 2010-2012 period at least 50% of professors achieved at least one publication indexed in the WoS. We censused all professors of Italian universities on staff in the whole 2001-2012 period in such SDSs (30,866 in all). We then divided this period into three-years intervals: such publication window ensures robust measures of performance (Abramo, D'Angelo & Cicero, 2012), and is in line with previous literature (He, Geng & Campbell-Hunt, 2009; Gonzalez-Brambila, Veloso & Krackhardt, 2013; Ductor, 2015).

The population used in our analysis was extracted from the Ministry of Education, Universities and Research (MIUR) database.[3] This database indexes names, academic rank, affiliation, and SDS of all professors in Italian universities at the end of each year. Next, the bibliometric publication dataset of these scientists was extracted from the Italian Observatory of Public Research (ORP), a database developed and maintained by the authors and derived under license from Thomson Reuters. Beginning from the raw data of 2001-2012 Italian publications in WoS, and applying a complex algorithm for disambiguation of the true identity of the authors and their institutional affiliations (for details see D'Angelo, Giuffrida & Abramo, 2011), each publication[4] is attributed to the university scientists that produced it, with a harmonic average of precision and recall (F-measure) equal to 96 (error of 4%).

Thanks to the assignment of each publication to each academic, we are able to identify all the productive academics with at least one publication in all the three-years periods under analysis.[5] The birth date of the academics is not provided in the MIUR database. We obtain it through analysis of the lists compiled by the MIUR beginning in 2004, showing the national academics with the right to vote in elections for members in the career-advancement committees. From these lists we identify the birth dates for

---

[2] The complete list is on http://attiministeriali.miur.it/UserFiles/115.htm, last accessed 4 September 2017.
[3] http://cercauniversita.cineca.it/php5/docenti/cerca.php, last accessed 4 September 2017.
[4] We exclude those document types that cannot be strictly considered as true research products, such as editorial material, conference abstracts, replies to letters, etc.
[5] The reason for this choice is the need to calculate the propensity to collaborate in the three-year periods under analysis, as illustrated in the next section of the paper.

16,823 academics, representing 54.5% of the professors active in all the periods under analysis. The missing data refer mainly to associate (coverage at 51.1%) and assistant professors (coverage at 37.2%). This is because since 2008, only full professors (coverage at 71.1%) are eligible to serve as members of the career-advancement committees. Table 1 shows that the level of coverage is not balanced across the different disciplines, with a minimum of 20.1% in Economics and Statistics and a maximum of 80.6% in Chemistry. This different level of coverage between disciplines could be due, other than to a different availability of data on birth dates, to different share of "unproductive" academics, given that scientists belonging to certain UDAs (Economics and statistics, Pedagogy and psychology, Civil engineering) tend to publish research output not only in journals censused by WoS, but also in other journals, conference papers and books that are sometimes only of national interest (Lariviere, Gingras & Archambault, 2006).

*Table 1: Distribution of academics in the dataset among UDAs*

| UDA | Total professors | Active in 2001-2003, 2004-2006, 2007-2009, 2010-2012 | Dataset | Coverage (%) |
|---|---|---|---|---|
| Mathematics and computer sciences (MAT) | 4,051 | 2,687 | 1,320 | 49.1 |
| Physics (PHY) | 3,374 | 2,068 | 1,566 | 75.7 |
| Chemistry (CHE) | 4,210 | 2,580 | 2,079 | 80.6 |
| Earth sciences (EAR) | 1,689 | 979 | 406 | 41.5 |
| Biology (BIO) | 6,787 | 4,125 | 2,803 | 68.0 |
| Medicine (MED) | 14,684 | 8,922 | 4,743 | 53.2 |
| Pedagogy and psychology (PPS) | 1,277 | 622 | 201 | 32.3 |
| Agricultural and veterinary sciences (AVS) | 3,646 | 2,296 | 989 | 43.1 |
| Civil engineering (CEN) | 2,008 | 1,230 | 348 | 28.3 |
| Industrial and information engineering (IIE) | 6,617 | 3,952 | 2,085 | 52.8 |
| Economics and statistics (ECS) | 2,304 | 1,405 | 283 | 20.1 |
| Total | 50,647 | 30,866 | 16,823 | 54.5 |

*3.2 Indicators of collaboration and research productivity*

One of the main limits of the literature on research collaborations is presented by the inability to identify all the types of collaboration that generated each publication (overall, intra-university, domestic, international) (He, Geng & Campbell-Hunt, 2009). We can take advantage on unequivocal identification of each academic with her/his home university (this operation is not possible for the non-academic co-authors),[6] so for each publication in the dataset, we have:

---

[6] Since the identification of the different types of collaboration is obtained by analysis of the authors and the addresses associated with each collaboration, our algorithm may not be capable of identifying all the intra-university collaborations, particularly those between a faculty member and other non-faculty colleague from the same university (for example, the collaborations between a full professor and their PhD students). However, these types of intramural collaborations do not appear relevant for the purposes of the proposed study.

- the complete list of all coauthors;
- the complete list of all their addresses;
- a sub-list of only the academic authors, with their SDS and university affiliations.

In this way, for each academic $i$ of the dataset, we measure the propensity to collaborate, by form of collaboration and overall. In detail, we measure the following indicators:

- Propensity to collaborate $C = \frac{cp_{i,t}}{N_{i,t}}$, where $cp_{i,t}$ is the number of publications resulting from collaborations over the period and $N_{i,t}$ is the total number of publications written by each academic over the period;
- Propensity to collaborate at the intra-university level $CI = \frac{cip_{i,t}}{N_{i,t}}$, where $cip_{i,t}$ is the number of publications resulting from collaborations with other academics belonging to the same university over the period;
- Propensity to collaborate extramurally at the domestic level $CED = \frac{cedp_{i,t}}{N_{i,t}}$, where $cedp_{i,t}$ is the number of publications resulting from collaborations with scientists belonging to other domestic organizations over the period;
- Propensity to collaborate extramurally at the international level $CEF = \frac{cefp_{i,t}}{N_{i,t}}$, where $cefp_{i,t}$ is the number of publications resulting from collaborations with scientists belonging to foreign organizations over the period.

These indicators vary between zero (if, in the observed period, the scientist under observation did not produce any publications resulting from the form of collaboration analyzed), and 1 (if the scientist produced all his/her publications through that form of collaboration).[7]

In order to calculate the research productivity of each professor we need to adopt a few simplifications and assumptions. Because data on production factors available to each professor are not available, we assume they are the same. Another assumption is that the hours devoted to research are more or less the same for all professors. Given the traits of the Italian academic system, the above assumptions appear acceptable.

The indicator of labor research productivity is measured through a proxy called Fractional Scientific Strength (*FSS*). In formula:

$$FSS = \frac{1}{t}\sum_{j=1}^{N}\frac{c_j}{\bar{c}}f_j$$

[1]

Where:
$t$ = number of years of work of the professor in period under observation
$N$ = number of publications of the professor in period under observation
$c_j$ = citations received by publication $j$
$\bar{c}$ = average of distribution of citations received for all cited publications in same year and subject category of publication $j$
$f_j$ = fractional contribution of the professor to publication $j$.

Fractional contribution equals the inverse of the number of authors, in those fields

---

[7] Similar indicators are presented by Martin-Sempere, Garzon-Garcia and Rey-Rocha (2008), Abramo, D'Angelo and Murgia (2013a) and Ductor (2015).

where the practice is to place the authors in simple alphabetical order.[8] Differently from other indicators of research performance, *FSS* embeds both quantity and impact of production. A thorough explanation of the theory and assumptions underlying *FSS* can be found in Abramo and D'Angelo (2014). For the purpose of this paper, given the varying publication and citation intensity between the SDSs under observation (Abramo & D'Angelo, 2015), we consider the rescaled value of productivity, or:

$$FSS_{scaled} = \frac{FSS}{FSS^*}$$

[2]

Where:

$FSS^*$ = average of productivity of all scientist operating in the same SDS.

We measure research productivity and the propensities to collaborate over the three-year periods 2001-2003, 2004-2006, 2007-2009, and 2010-2012. In practice, we divide the period of observation into four partitions, each one representing a triennium indicated with the variable *t*. In line with the previous literature (He, Geng & Campbell-Hunt, 2009; Gonzalez-Brambila, Veloso & Krackhardt, 2013; Ductor, 2015), we choose this long time period for the evaluation of collaboration and research productivity, above all because some time is needed before the collaboration achieves concrete form in the publications. Moreover, in this manner we ensure satisfactory levels of reliability in our analyses (Abramo, D'Angelo & Cicero, 2012), since the lengthy time periods reduce the impact of the delays in the publication process, which affect some of the research fields more than others. Table 2 presents descriptive statistics of dependent and independent variables of the analysis (the same statistics at discipline level are presented in Table S.1 of Supplementary material). In line with Abramo, D'Angelo and Murgia (2013a), the average value of *C* is very close to 1, seeing as almost all the scientific production features the occurrence of co-authorship, while the propensity for the individual types of collaboration show average values of 0.74 (*CI*), 0.49 (*CED*) and 0.25 (*CEF*). Concerning the demographic variables, the dataset consists of just over 73% men, while the average birth year is 1955. In this paper the cohort is defined by the birth year. The use of cohort rather than age is due to the adoption of a longitudinal approach, which makes it difficult to disentangle the effect of age and calendar year (Levin & Stephan, 1991). Concerning the academic variables, these are measured as of the day before the beginning of each time interval (31/12/2000, 31/12/2003, 31/12/2006 and 31/12/2009). At the first time interval, the dataset consists of 26% full professors, 27% associates, 33% assistant professors, and 14% adjunct professors. At the last stage, the dataset consists of 45% full professors, 33% associates and 22% assistant professors. We associate the academic rank to an ordinal variable varying from 1 (adjunct professor) to 4 (full professor).

---

[8] For the life sciences, widespread practice in Italy and abroad is for the authors to indicate the various contributions to the published research by the order of the names in the byline. For these areas, we give different weights to each co-author according to their order in the byline and the character of the co-authorship (intramural or extra-mural). If first and last authors belong to the same university, 40% of citations are attributed to each of them; the remaining 20% are divided among all other authors. If the first two and last two authors belong to different universities, 30% of citations are attributed to first and last authors; 15% of citations are attributed to second and last but one author; the remaining 10% are divided among all others. The weighting values were assigned following advice from senior Italian professors in the life sciences. The values could be changed to suit different practices in other national contexts.

*Table 2: Descriptive statistics of dependent and independent variables (16,823 observations)*

| Variable | Mean | Std dev. |
|---|---|---|
| Fractional Scientific Strength *FSS* | 1.18 | 1.84 |
| Propensity to collaborate *C* | 0.97 | 0.11 |
| Propensity to collaborate within university *CI* | 0.74 | 0.32 |
| Propensity to collaborate extramurally at the domestic level *CED* | 0.49 | 0.34 |
| Propensity to collaborate extramurally at the international level *CEF* | 0.25 | 0.28 |
| Rank (1 = Adjunct, 2 = Assistant, 3 = Associate, 4 = Full professor) | 3.02 | 0.89 |
| Cohort (Birth year) | 1955.27 | 8.59 |
| Gender (Male) | 0.73 | 0.44 |

## 3.3 Cross-lagged panel models

To analyze the relation between the research productivity, the propensity to the different forms of collaboration and the various determinants considered, we have to solve the following dynamic equations:

$$FSS_{it} = \alpha_i + \beta_1 FSS_{it-1} + \beta_2 C_{it-1} + \beta_3 CI_{it-1} + \beta_4 CED_{it-1} + \beta_5 CEF_{it-1} + \beta_6 R_{it-1} + \beta_7 V_{it-1} + \varphi t + \varepsilon_{it} \quad [3]$$

$$Y_{it} = \mu_i + \delta_1 Y_{it-1} + \delta_2 FSS_{it-1} + \delta_3 R_{it-1} + \delta_4 V_{it-1} + \tau t + \xi_{it} \quad [4]$$

where the equation [4] has been repeated for each one of the four indicators of propensity to collaborate ($Y = C, CI, CED, CEF$). $\varepsilon_{it}$ and $\xi_{it}$ represent pure random noise, while $\alpha_i$ and $\mu_i$ represent individual-specific unobserved heterogeneity in *FSS* and *Y*, supposed constant over time. So, $\alpha_i$ accounts for the fact that research productivity of a scientist could be affected by unobserved variables, such as his/her scientific ability or intrinsic motivation, that distinguish him/her from the other academics. Similarly, $\mu_i$ accounts for the fact that a scientist's propensity to collaborate could be affected by other unobserved variables, such as his/her extroversion or language proficiency, that distinguish him/her from the other academics. The specification of individual-specific unobserved effects helps to separate the within-person level from the between-person level, improving the analysis of a process that takes place at the within-person level (Hamaker, Kuiper & Grasman, 2015). In line with Allison (2005) and Teachman, Duncan, Yeung and Levy (2001), we model these effects by allowing for them to be correlated with all time-varying exogenous predictors, while we assume that they are uncorrelated with time-invariant variable *V* represents the set of independent variables like cohort and gender.[9] We add *t* as predictor, in order to partially control for the impact of unobserved variables that could homogeneously affect the academics in their research productivity and propensity to collaborate, such as enacting relevant national policies. Regarding the academic rank (*R*) of a scientist, because it should depend on his/her past research productivity and propensity to collaborate, we cannot consider it as a strictly exogenous predictor. Therefore, we measure the dynamics of academic rank as function of the past research productivity and propensity to collaborate, thanks to the following equation:

$$R_{it} = \eta_i + \theta_1 R_{it-1} + \theta_2 FSS_{it-1} + \theta_3 FSS_{it-2} + \theta_4 C_{it-1} + \theta_5 CI_{it-1} + \theta_6 CED_{it-1} + \theta_7 CEF_{it-1} + \theta_8 V_{it-1} + \kappa t + v_{it} \quad [5]$$

where $v_{it}$ and $\eta_i$ represent, respectively, pure random noise and individual-specific unobserved heterogeneity. In this equation, we include also the impact of two-lagged research productivity because, in the Italian university system, promotions seem to be based on the evaluation of long-term, rather than short-term performance (Abramo, D'Angelo & Murgia, 2016).

---

[9] As discussed before, the reference case for this dummy variable is Gender = female.

The decision to adopt such dynamic models, with the lagged dependent variable as additional explanatory variable, allows to control for the within-person carry over effect (Hamaker, Kuiper & Grasman, 2015; Kuppens, Allen & Sheeber, 2010). Our indicators of research productivity and collaboration are computed on a three-year interval, so that each lagged variable is computed on the three years preceding the period under analysis. Similarly, as discussed above, academic rank is observed the day before the beginning of each time interval. In this way, the use of lagged variables as predictors allows to detect the relationship among variables, avoiding simultaneity between research productivity, collaboration, and academic rank.

To estimate the dynamic equations [3], [4] and [5] we test different cross-lagged panel models (Finkel, 1995; Little, Preacher, Selig & Card, 2007) by using Structural Equation Modeling (SEM). Even if SEM has been developed especially in psychological and sociological literature (Finkel, 1995), it has been gaining popularity even in economic and management literature (Shin & Konrad, 2017; Protogerou, Caloghirou & Lioukas, 2012). Differently from other statistical approaches, SEM deals with the endogeneity among research productivity, propensities to collaborate and their determinants, thanks to the simultaneous estimation of the equations [3], [4] and [5]. We create different recursive models based on these equations, guaranteeing the respect of the three principles of causality suggested by Gollob and Reichardt (1991), i.e. the ordering of causes and outcomes, the control of autoregressive influences, and the use of an adequate time lag length. In particular, in line with Wooldridge (2002) and Allison (2005), we model the assumption of sequential exogeneity for the dependent variables in the equations [3], [4] and [5], i.e. *FSS*, *Y* and *R*. This means that research productivity, propensities to collaboration, and academic rank are independent of all the future values of their pure random noises, but may be correlated with the past values of their pure random noises. In this way, we allow for an unobserved factor affecting performance at time *t*, such as a sick leave, to impact only the values of the dependent variables time *t+1* onward. Differently, we assume that all the other independent variables are strictly exogenous and that all pure random noises are uncorrelated.[10] By solving a system among these equations, SEM allows to estimate different cross-lagged panel models, whose main structure is presented in Figure 2.[11] In line with the equations [3], [4] and [5], Figure 2 shows the relationships among research productivity, propensity to collaborate at international level and academic rank, in three time periods. The dashed lines from propensity to collaborate to research productivity represent the relationship hypothesized in *H1*, while the dotted lines from research productivity and propensity to collaborate represent the relationship hypothesized in *H2*. In particular, we test four different cross-lagged panel models, based on the following structure:

- Model A, without the dashed and the dotted lines (both *H1* and *H2* not considered);
- Model B, with the dashed, but not the dotted lines (*H1* considered, *H2* not considered);
- Model C, with the dotted, but not the dashed lines (*H1* not considered, *H2* considered);
- Model D, with the dashed and the dotted lines (both *H1* and *H2* considered).

---

[10] Our SEM models aim at realistically describing the relationship among research productivity, collaboration and their determinants, but taking into consideration also the problems related to the computation of these models. We are aware that a more realistic analysis would require adding or modifying some assumptions of our SEM models. For example, our assumption of no common omitted causes in the research productivity and collaboration models could be considered as restrictive and unrealistic. Only if the assumptions realistically hold, SEM models could provide a correct explanation of the causal relationships among the variables under study. Differently, the relationships among the variables could be interpreted as causal relationships *a la Granger*.

[11] For reasons of clarity, we avoid to present in the Figure 2 the whole SEM models estimated, which are characterized by the presence of 4 time periods, of different forms of propensity to collaborate that work simultaneously, of time invariant variables like gender and birth year, of time dummies, of pure random noises and individual-specific unobserved effects.

We evaluate each model by using the measures of goodness of fit traditionally adopted in SEM literature. Besides, we evaluate them on the base of Chi-squared pair comparisons, aimed at detecting the model that better fits the original data.

*Figure 2: Main structure of the cross-lagged panel models under analysis*

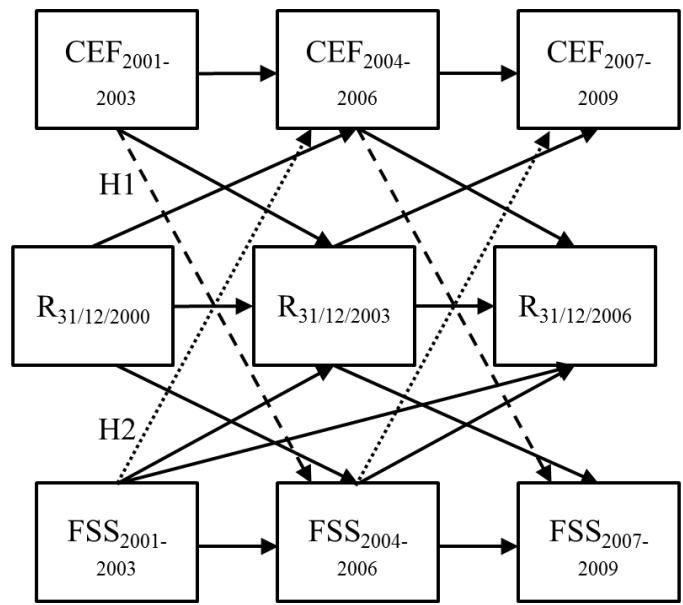

Since we assume the time invariance of the predictors' effect on the dependent variable, we develop all the models by imposing several equality constraints. In any case, we evaluate the goodness of this choice by computing the Lagrange multiplier test indices related to each equality constraint.

In order to deal with the potential heteroscedasticity of our data, we estimate our models by using the two-stage robust method developed by Yuan and Zhong (2008) and implemented in the SAS procedure CALIS. This two-stage robust method estimates robust covariance and mean matrices in the first stage, and then feeds the robust covariance and mean matrices for ML estimation in the second stage. In this way, for each regression, we obtain coefficients' standard errors and the level of variance explained by each equation, by presenting R-squared for OLS models. In all the regressions executed, we have also checked for the absence of multi-collinearity between the explanatory variables, then omitting few of them as appropriate.

Thanks to the estimation of these cross-lagged panel models, we are also able to compute the value of the indirect effect of a predictor on a dependent variable. The indirect effect of an independent variable ("treatment") to a dependent variable ("outcome") is due to the presence of one or more variables ("mediators"), which are influenced by the treatment, and in turn influence the value of the outcome. In this work we report the value of the indirect effect, and its ratio to the indirect effect regarding the same variables.[12] Because in cross-lagged panel models the effects between variables propagate over time, in line with Selig and Preacher (2009) we evaluate the indirect effect on an outcome at time $t$ by considering the treatment at $t - 2$. So we evaluate the indirect effect of rank (gender or cohort) at $t - 2$ on research productivity measured at time $t$, and the indirect effect of rank (gender or cohort) at $t - 2$ on the propensities to collaborate measured at time $t$.

## 4. Results

Using the indicators and analytical methods discussed in the previous section, we now analyze the relations between research productivity, the propensities to the different forms of collaboration, and the personal variables considered, beginning from the academic rank. In Table 3, we present the correlation matrix among the variables under analysis. Given the large number of observations, almost all the relationships between the variables appear to be significant. In absolute value, the highest correlation is between academic rank and cohort, evaluated in term of birth year. This result is in line with the relationship between cohort and probability to promotion highlighted by Pezzoni, Sterzi and Lissoni (2012). Among the different forms of propensity to collaborate, we find a negative correlation between propensity to collaborate within university with propensity to collaborate extramurally both at the domestic level and at the international level. This result seems to reveal that the collaborations with other academics of the same university avoid the need for collaborations at extra-mural level, and vice versa. In any case, the highest correlation between different forms of propensity to collaborate is between the collaboration in general and within university. This result is due to the fact that collaboration in general results from the sum of all the specific forms of collaboration, and collaboration within university, as shown in Table 2, is the most diffused. Given the scarce additional information provided by $C$, the high level of correlation between this variable and the other propensities to collaborate, and the fact that the low variance of this indicator (0.12) negatively affects the quality of the estimations[13], we omit this variable from the following analyses.

To test the hypotheses *H1* and *H2*, we first calculate the four cross-lagged panel models (Models 1-4) described in the previous section. The upper section of Table 4 presents the results concerning the determinants of research productivity [eq. 3], the central section presents the results for the determinants of the different forms of propensity to collaborate [eq. 4], while the lower section presents the results for the determinants of academic rank [eq. 5]. Finally, the last rows of the table present different measures of overall goodness of fit that allow to evaluate each model and compare them. Apart from Chi-squared, which is sensible to the sample size (Bagozzi & Yi, 1988), all the indicators show that the four competing models provide a good overall fit. In order to detect which model provides the best overall fit, we compare each pair of models, based on their Chi-squared and the related degrees of freedom. Given that diff($\chi^2$)$_{A-D}$= 3609.650 (p = 0.0000); diff($\chi^2$)$_{B-D}$= 8.460 (p = 0.0374); and diff($\chi^2$)$_{C-D}$= 18.909 (p =0.0003), Model D provides the best overall fit among the four models under analysis, as confirmed also by other goodness of fit indicators, such as Akaike Information Criterion (AIC), Standardized Root Mean Square Residual (SRMR) and Root Mean Square Error of Approximation (RMSEA). This model contains both relationships, from *FSS* to propensity to collaborate and vice versa, related to *H1* and *H2*. In the following, we discuss the results related to these hypotheses, and to the relationships among the other variables, by taking into consideration the estimations of Model D, but indicating the relevant differences with the other models estimated.

---

[12] To assess the significance of the indirect effect, we should apply the Sobel test (Sobel, 1982) or, better, a more robust version of it, based on bootstrapping (Preacher & Hayes, 2008). Nevertheless, in our cross-lagged panel models, the computation of the standard error and the significance level of the indirect effect is made very complicated by the fact that there are many parallel and serial mediators, in addition to the assumption of weak exogeneity.

[13] The SEM estimation of the models with C shows some problems in the optimization process. In any case, the differences with the models presented in the paper are very limited. The authors would be pleased to provide these results on request.

*Table 3: Correlation matrix (16,823 observations)*

|  | 1 | 2 | 3 | 4 | 5 | 6 | 7 |
|---|---|---|---|---|---|---|---|
| 1. FSS |  |  |  |  |  |  |  |
| 2. C | -0.041*** |  |  |  |  |  |  |
| 3. CI | -0.115*** | 0.338*** |  |  |  |  |  |
| 4. CED | -0.012** | 0.192*** | -0.261*** |  |  |  |  |
| 5. CEF | 0.103*** | 0.083*** | -0.262*** | 0.135*** |  |  |  |
| 6. Rank | 0.136*** | 0.019*** | -0.051*** | 0.008* | 0.045*** |  |  |
| 7. Cohort | 0.017*** | -0.064*** | -0.049*** | -0.082*** | 0.012** | -0.565*** |  |
| 8. Gender | 0.103*** | -0.043*** | -0.083*** | -0.021*** | 0.019*** | 0.235*** | -0.157*** |

*Significance level:* ***= $p < 0.001$; **= $p < 0.01$; *= $p < 0.05$

*Table 4: Cross-lagged panel model results (16,823 observations)*

| FSS | Model A | Model B | Model C | Model D |
|---|---|---|---|---|
| Intercept | -26.509*** (1.753) | -25.392 (64.277) | -26.509*** (1.753) | -25.638 (65.652) |
| FSS | 0.312*** (0.009) | 0.319*** (0.009) | 0.312*** (0.009) | 0.319*** (0.009) |
| CI |  | 0.073** (0.025) |  | 0.102*** (0.026) |
| CED |  | 0.058** (0.020) |  | 0.081*** (0.022) |
| CEF |  | -0.048° (0.025) |  | -0.029 (0.026) |
| Rank | 0.018 (0.012) | 0.017 (0.013) | 0.018 (0.012) | 0.017 (0.013) |
| Cohort | 0.014*** (0.001) | 0.013 (0.032) | 0.014*** (0.001) | 0.014 (0.033) |
| Gender (Male) | 0.288** (0.107) | -0.473 (1.550) | 0.288** (0.107) | -0.433 (1.583) |
| R-squared (%) | 58.01 | 59.60 | 58.01 | 59.56 |
| **CI** | | | | |
| Intercept | 22.675*** (0.002) | 22.675*** (0.002) | 22.999*** (2.413) | 22.954*** (1.896) |
| FSS |  |  | 0.003* (0.001) | 0.005*** (0.002) |
| CI | 0.318*** (0.009) | 0.318*** (0.009) | 0.311*** (0.009) | 0.311*** (0.009) |
| Rank | 0.005° (0.003) | 0.005° (0.003) | 0.006° (0.003) | 0.006° (0.003) |
| Cohort | -0.012*** (0.000) | -0.012*** (0.000) | -0.012*** (0.001) | -0.011*** (0.001) |
| Gender (Male) | 0.627*** (0.013) | 0.627*** (0.013) | 0.493*** (0.057) | -0.266*** (0.045) |
| R-squared (%) | 52.12 | 52.12 | 51.15 | 51.12 |
| **CED** | | | | |
| Intercept | 21.205*** (0.002) | 21.205*** (0.002) | 21.449*** (3.332) | 21.421*** (3.271) |
| FSS |  |  | 0.001 (0.002) | 0.002 (0.002) |
| CED | 0.255*** (0.008) | 0.255*** (0.008) | 0.249*** (0.008) | 0.249*** (0.008) |
| Rank | 0.003 (0.004) | 0.003 (0.004) | 0.003 (0.004) | 0.003 (0.004) |
| Cohort | -0.011*** (0.000) | -0.011*** (0.000) | -0.011*** (0.002) | -0.010*** (0.002) |
| Gender (Male) | 0.429*** (0.012) | 0.429*** (0.012) | 0.887*** (0.079) | -0.952*** (0.077) |
| R-squared (%) | 48.27 | 48.27 | 47.75 | 47.74 |
| **CEF** | | | | |
| Intercept | 20.351*** (0.002) | 20.351*** (0.002) | 20.85*** (1.795) | 20.617*** (1.986) |
| FSS |  |  | 0.004** (0.001) | 0.003* (0.001) |
| CEF | 0.205*** (0.009) | 0.205*** (0.009) | 0.200*** (0.009) | 0.200*** (0.009) |
| Rank | 0.003 (0.003) | 0.003 (0.003) | 0.003 (0.003) | 0.003 (0.003) |
| Cohort | -0.010*** (0.000) | -0.010*** (0.000) | -0.011*** (0.001) | -0.010*** (0.001) |
| Gender (Male) | -0.271*** (0.009) | -0.271*** (0.009) | -0.082° (0.043) | -0.312*** (0.047) |
| R-squared (%) | 46.71 | 46.71 | 45.88 | 45.88 |
| **Rank** | | | | |
| Intercept | -47.412*** (2.209) | -47.412*** (2.209) | -47.412*** (2.209) | -47.412*** (2.209) |
| FSS | 0.007* (0.003) | 0.007* (0.003) | 0.007* (0.003) | 0.007* (0.003) |
| FSS-2 | 0.009*** (0.003) | 0.009*** (0.003) | 0.009*** (0.003) | 0.009*** (0.003) |
| CI | 0.000 (0.012) | 0.000 (0.012) | 0.000 (0.012) | 0.000 (0.012) |
| CED | 0.000 (0.01) | 0.000 (0.01) | 0.000 (0.01) | 0.000 (0.01) |
| CEF | 0.015 (0.013) | 0.015 (0.013) | 0.015 (0.013) | 0.015 (0.013) |
| Rank | 0.215*** (0.009) | 0.215*** (0.009) | 0.215*** (0.009) | 0.215*** (0.009) |
| Cohort | 0.026*** (0.001) | 0.026*** (0.001) | 0.026*** (0.001) | 0.026*** (0.001) |
| Gender (Male) | -0.679*** (0.017) | -0.679*** (0.017) | -0.679*** (0.017) | -0.679*** (0.017) |
| R-squared (%) | 87.39 | 87.39 | 87.39 | 87.39 |
| $\chi^2$ | 16617.517 | 13016.327 | 13026.776 | 13007.867 |
| d.f. | 148 | 145 | 145 | 142 |
| Prob. > $\chi^2$ | 0.0001 | 0.0001 | 0.0001 | 0.0001 |
| SRMR | 0.170 | 0.026 | 0.026 | 0.026 |
| RMSEA | 0.081 | 0.073 | 0.073 | 0.073 |
| CFI | 0.947 | 0.958 | 0.958 | 0.958 |
| TLI | 0.917 | 0.934 | 0.934 | 0.932 |
| AIC | 16871.517 | 13276.327 | 13286.776 | 13273.867 |

*Standard errors in brackets. Significance level: \*\*\*=p < 0.001; \*\*= p < 0.01; \*=p < 0.05; ° = p < 0.1*

The analysis of the results concerning the determinants of research productivity and propensities to collaborate show the high impact of the same variable measured in the preceding three years. This result reveals the presence of a positive within-person carry over effect in the phenomena related to these variables, so that a scientist with good research productivity or propensity to collaborate at time $t$ is likely to gain another good performance at time $t+1$, and vice versa (Hamaker, Kuiper & Grasman, 2015; Kuppens, Allen & Sheeber, 2010). Among different kind of propensities to collaborate, we note that the within-person carry over effect is higher at intra-university level and lower at the international level. These results could be related to the different costs of maintaining international collaborations over time. In general, the carry-over effect on research productivity and on propensities to collaborate occurs also at discipline level (results are shown in Table S.2 of Supplementary material). The carry-over effect is always positive and significant, but in Pedagogy and psychology, whereby the effect is positive, but unsignificanton research productivity and on the propensity to collaborate at domestic level.

Analyzing the results concerning the impact of the different forms of propensity to collaborate on research productivity, we note that only the propensities to collaborate at the intra-university (*CI*) and domestic (*CED*) levels have a positive and significant impact on research productivity, while the propensity to collaborate at international level (*CEF*) has a negative, but not significant, impact on research productivity. Findings partially confirmed at discipline level, In fact, the propensity to collaborate at the intra-university (*CI*) has a significant and positive effect in Industrial and information engineering, while at domestic level (*CED*) is never significant. The propensity to collaborate at the international level (*CEF*) is never significant either but, differently from the overall level, its effect is positive in Biology, Mathematics and computer sciences, and Pedagogy and psychology. Even if the positive impact of collaboration on research productivity is in line with *H1*, contrary to this hypothesis, our results show that collaboration at the intra-university and domestic levels have a more positive impact than collaboration at international level. These results could be explained by taking into consideration the lower costs involved in the activation of such collaborations, the easier transmission of tacit knowledge, and the more simple creation of a climate of trust (He, Geng & Campbell-Hunt, 2009; Smeby & Try, 2005). The negative result for the propensity to collaborate at the international level, which contrasts with the literature (Lissoni, Mairesse, Montobbio & Pezzoni, 2011; He, Geng & Campbell-Hunt, 2009; Smeby & Try, 2005), reveals the general inability of Italian academics to take advantage of the potential gains from this form of collaboration, such as better visibility, and the access to complementary resources and backgrounds. This could be due, other than to the natural higher cost of this form of collaboration, to the inability of the academics to fully take advantage of the new communication systems that could allow a better coordination among scientists from different countries. This explanation is aligned with Hamermesh and Oster (2002), who show that publications co-authored by far away institutions are of lower quality than those from closer institutions. Hamermesh and Oster (2002) explain their findings, stating that scientific collaborations do not respond only to the authors' need to increase their productivity, but also the consumption value stemming from the interaction with colleagues. Collaborations from a far distance must have a higher consumption value, due to the willingness to keep a relationship often started in the same institution in the early stage of individuals' career.

The results concerning the impact of research productivity on the different forms of propensity to collaborate generally support *H2*, even if the positive impact of research productivity is significant only for propensity to collaborate at intramural and international level. Research productivity then seems to favor the development of new collaborations, thanks to higher visibility and better ability in their management (Abramo, D'Angelo & Solazzi, 2011; Kato & Ando, 2013). Results are only partially confirmed at discipline level: research productivity shows a significant and positive impact on the propensity to collaborate at intramural level solely in Civil engineering, Industrial and information engineering and Chemistry; at international level only in Earth sciences and Medicine. Differently from the overall level, at domestic level only in Chemistry.

Because Model B excludes the relationships related to *H2* and Model C excludes the relationships related to *H1*, it is useful to compare them with Model D, in order to evaluate the robustness of the results related to both these hypotheses. This comparison shows that the coefficients related to the impact of research productivity on the different forms of research collaboration vary in a very limited manner in Model C and D. Similarly, the coefficients related to the impact of the different forms of research collaboration on research productivity vary in a limited manner in Model B and D.

The analysis of academic rank[14] as predictor of research productivity and collaboration shows a weak support for *H3*. Indeed, the positive impact of academic rank on research productivity is not significant, while academic rank shows a positive and weakly significant effect on the three forms of research collaboration, especially at intra-mural level. This latter result aligns with the large part of the previous literature, which highlight that full professors have the highest ability in the development of research collaborations (Melkers & Kiopa, 2010; Bozeman & Corley, 2004). In general results do not change at discipline level. These results are further confirmed by the analysis of the indirect effect of academic rank, presented in Table 5. This table shows that the indirect effect of academic rank on research productivity and on the three forms of research collaboration is positive. All these indirect effects account for about one half of the direct effect of academic rank on the same variables.

*Table 5: Indirect effects in Model D (16,823 observations)*

|  |  | FSS | CI | CED | CEF |
|---|---|---|---|---|---|
| Rank | Indirect effect | 0.010 | 0.003 | 0.002 | 0.001 |
|  | Indirect/Direct effect (%) | 57.9 | 54.2 | 47.7 | 43.6 |
| Cohort | Indirect effect | 0.003 | -0.003 | -0.002 | -0.002 |
|  | Indirect/Direct effect (%) | 22.6 | 29.2 | 23.8 | 18.8 |
| Gender | Indirect effect | -0.245 | -0.089 | -0.241 | -0.066 |
|  | Indirect/Direct effect (%) | 56.6 | 33.4 | 25.3 | 21.0 |

The cohort variable shows a positive but not significant effect on research productivity, while a negative and significant effect on all the propensities to collaboration. So, younger cohorts seem to perform better, but they have a lower propensity to collaborate than older ones. The results is confirmed by the analysis of the indirect effect, which accounts for less than one third of the direct effect on the same variables. Results are confirmed at discipline level, with the only exceptions of Earth sciences and Agricultural and veterinary sciences, whereby the cohort has a positive and significant effect on the propensity to collaboration at domestic level. The positive but week impact on research productivity is in line with *H4* This result could be explained by taking into consideration the greater risk of knowledge obsolescence for older cohorts and the evolution of university job market, with an increasing emphasis on publications (Levin & Stephan, 1991). Differently, the significant and negative effect of the cohort on the propensities to collaborate contrasts with *H4*. The reason for that could be the higher cumulative advantage of higher academic ranks, which make them more attractive to potential collaborators, and effective collaborations more likely (Abramo, D'Angelo & Murgia, 2014).

---

[14] Our focus is on the relationships among research productivity, collaboration and their determinants, and we introduce the equation [5] with the only intent to correctly model endogeneity of academic rank. We avoid to devote too much space on the the determinants of academic rank, so we briefly discuss the results related to equation [5] in the present note. In line with the previous literature (Lissoni, Mairesse, Montobbio & Pezzoni, 2011; Pezzoni, Sterzi & Lissoni, 2012), promotion is positively and significantly related to research productivity, especially that evaluated in the medium term (Abramo, D'Angelo & Murgia, 2016). Even propensity to collaborate at international level shows a positive but not significant relation with promotion (van Rijnsoever, Hessels & Vandeberg, 2008). Finally, promotion is significantly related to female and younger cohort of academics. We will analyze the determinants of academic rank in a future paper, specifically dedicated to this topic.

The results concerning the impact of gender on research collaboration contrasts with *H5* and with the large part of the preceding literature, which has indicated how female researchers are characterized by a higher difficulty to manage collaborations, especially at international level (Abramo, D'Angelo & Murgia, 2013b), and a lower research performance (Abramo, D'Angelo & Caprasecca, 2009; Kyvik & Teigen, 1996). Findings are confirmed by the analysis of the indirect effect, which accounts for more than one half of the direct effect on research productivity and for about one third of the direct effect on propensities to collaboration. At discipline level, the results regarding the impact of gender on research productivity are aligned with the overall level, while the impact on the propensities to collaborate notably varies across disciplines. For example, the impact of gender on propensity to collaborate at intramural level is positive and significant in Medicine, while at domestic and international level is positive and significant in Agricultural and veterinary sciences, and in Biology. The impact of gender on research productivity, even if unsignificant, could be attributed to the vanishing of gender gap, suggested recently by van Arensbergen, van der Weijden and van den Besselaar (2012). Similarly, the impact of gender on propensities to collaborate confirm that the obstacles for females to collaboration are on the wane (Abramo, D'Angelo & Murgia, 2013b). Nevertheless, it is important to highlight that our study analyzes the determinants of the within-person variation in research productivity. Indeed, this result suggests that female researchers have higher positive variation in research productivity and in propensities to collaboration, not that their absolute research productivity and propensities to collaboration are higher than those of their male colleagues. It is interesting to note that the impact of gender on research productivity and propensities to collaboration varies strongly from models D to the other models. These contrasting results could be explained by taking into consideration that only model D assesses at the same time the effect of research collaboration on productivity, and vice versa. The relations between research collaboration and research productivity require a complete model, for a correct estimate of the effect of gender on these variables.

## 5. Conclusions

The enormous growth in coauthored publications which has been registered in recent years has been favored by a variety of targeted policies on the part of the individual universities and national and international research systems. These policies, beginning from those concerning the financing of research projects, have often been motivated by the supposed benefits of collaboration (Lee & Bozeman, 2005; He, Geng & Campbell-Hunt, 2009). However, the literature of studies concerning the theme has not yet permitted full clarification of the causal links between collaboration, research productivity and some personal and organizational variables that can in fact influence both of these dimensions.

The present work contributes to further clarify this topic by means of a series of analyses on an extremely ample population of Italian academics, using an indicator of research productivity (*FSS*) that has thus far been applied only in Italy. Thanks to a cross-lagged panel model, we analyzed the impact of collaboration on research productivity, and vice versa. While the former relationship has been deeply analyzed by the literature, only few papers test the latter one. The analysis of both relationships allows a better understanding of the contextual processes of development of the scientists' scientific and technical human capital, and of their social capital (Bozeman & Corley, 2004).

In particular, our work fills a gap in the literature on the determinants of collaboration by showing that research productivity positively influences collaborations at intramural and international level. This fact could be explained by the attraction exercised by the most productive scientists, and by their greater ability in effectively managing collaborations.

Contrary to the literature, we show that only collaborations at domestic level have a positive impact on research productivity. This result could be explained by taking into consideration the higher costs related to international collaborations, which seem to overcome the benefits of this form of collaboration. In this sense, it will be useful to extend the analyses to other countries, in order to understand if this result could be generalized or is specifically related to university systems, like the Italian one, affected by structural barriers hindering international collaborations.

Another result that should be confirmed by similar analyses on other university systems is the one concerning the impact of academic rank. In fact, while its impact on collaboration is in line with the literature, the lack of significance of its effect on research productivity instills doubt on the use of this variable as a proxy of the scientist's cumulative advantage (Abramo, D'Angelo & Murgia, 2016; Kelchtermans & Veugelers, 2011). This hypothesis is based on the fact, confirmed even by the current analysis, that research productivity has a strong impact on promotion. So, higher ranks should be related to higher previous performance, which could favor an ever increasing productivity in the future. Eventually, the weak impact of rank on research productivity could be explained as the result of a past hiring and promotion system not adequately linked to productivity, and/or of a loss of motivation after the appointment to a higher academic rank.

Other than a re-examination of the theories on the relationship among research collaboration, productivity and their determinants, the present paper offers some useful insights for the development of policies related to these issues. First, these results confirm that the definition of policy intended to maximize research productivity must take due account of the question of the different types of collaboration, evaluating their benefits and costs. Second, the results suggest the necessity of policy that reinforces the capacity to collaborate in effective manner, above all on the part of academics of lower academic rank, and of older cohorts.

The robustness of our results is related to the assumptions in the model; an approach based on natural experiments could lead to more robust and definitive results on the relation between research productivity and collaboration. Azoulay, Graff Zivin and Wang (2010) and Waldinger (2010) adopted the latter approach to assess the impact of collaboration with specific scientists on research performance. However, conducting a natural experiment aimed at assessing the overall collaboration of a scientist is a formidable challenge.

To further assess the robustness of our results, the indirect effect of the analysed variables, we should adopt a sensitivity analysis, which could permit the quantification of the impact of any potential omitted confounders. In this sense, the development of more advanced software packages for causal inference approach (Imai, Keele & Tingley, 2010) will permit such sensitivity analysis. In future, we could attempt to broaden our dataset with information serving precisely to map some of these confounders, for example the individual's fluency in one or more foreign languages, which could influence both the research productivity and the propensity to collaborate. Future research could delve into the impact of interdisciplinary collaboration, or of collaboration between academics belonging to different cohorts.